# SuperDense Teleportation using Hyperentangled Photons

Trent M. Graham[1], Herbert J. Bernstein[2], Tzu-Chieh Wei[3], Marius Junge[4], & Paul G. Kwiat[1]

[1]*Department of Physics, University of Illinois at Urbana-Champaign, 1110 W. Green Street, Urbana, IL 61801, USA*
[2]*School of Natural Science and ISIS Institute for Science and Interdisciplinary Studies, Hampshire College, Amherst, MA 01002, USA*
[3]*C.N. Yang Institute for Theoretical Physics and Department of Physics and Astronomy, State University of New York at Stony Brook, Stony Brook, NY, USA*
[4]*Department of Mathematics, University of Illinois at Urbana-Champaign, 1409 W. Green Street, Urbana, IL 61801, USA*
Author e-mail address: tgraham2@illinois.edu

**ABSTRACT**

We use photon pairs hyperentangled in polarization and orbital angular momentum to implement a novel entanglement-enhanced quantum state communication technique, known as SuperDense Teleportation, to communicate a specific class of single-photon ququart states between two remote parties, with an average fidelity of 87.0(1)%, almost twice the classical limit of 44%. We compare this technique with quantum teleportation and remote state preparation and show that SuperDense teleportation requires less classical information and fewer experimental resources. We discuss the information content of this constrained set of states and demonstrate that this set has an exponentially larger state space volume than the lower dimensional general states with the same number of state parameters.

## Introduction

The transfer of quantum information over long distances has long been a goal of quantum information science. Loss is particularly devastating to quantum communication channels as quantum states cannot be amplified [1]. Moreover, random fluctuations in the communication channel can reduce the coherence of a quantum state, and error correction protocols for quantum states are presently very difficult to implement in practice [2]. However, if the sender (Alice) and the receiver (Bob) already share an entangled pair of qubits, then they may use a number of techniques to transfer quantum states using only classical information channels. In single-qubit quantum teleportation (QT, Fig. 1a) [3], Alice performs a measurement in the Bell state (i.e., maximally entangled) basis on the unknown state provided by a state chooser (Charles) and her half of the entangled state that she shares with Bob. She then sends the two-bit outcome of her measurement to Bob over a classical communication channel. Based on Alice's message, Bob performs one of four unitary transformations on his half of the originally entangled pair, transforming it into the exact state that Charles chose. Teleportation has been successfully demonstrated with probabilistic protocols for photons [4,5,6,7] and with deterministic protocols using nonlinear interactions for ions, atoms, superconducting qubits, and hybrid systems between photons and ions [8,9,10,11,12]. More recently, QT has been performed using photons entangled in spatial mode where Charles' quantum state is encoded on the polarization degree of freedom of Alice's photon [13]. Since Bell measurements between photonic degrees of freedom do not require nonlinear interactions, this protocol could theoretically be implemented with 100% efficiency.

In teleportation, Charles provides a quantum state which he wishes to be sent to Bob. However, if Charles is instead allowed to encode his desired state parameters he wishes to send directly on Alice's half of the entangled state, then a simpler method may be used to transmit the unknown qubit state from Alice to Bob. In this technique, known as remote state preparation (RSP, Fig. 1b), Alice need only perform measurements on a single qubit and transmit the outcome to Bob [14]. Then, as in teleportation, Bob performs a unitary transformation on his qubit, based on the message he received. It might be speculated, since Alice performs her measurement only on a single-qubit state, that she would only have to send a single-bit message to Bob. However, because Bob cannot perform a universal NOT gate (a mapping of the input state to its orthogonal) [15], a one-bit message from Alice is generally not sufficient for him to convert his state to the one Charles wished to send [14]. In fact, because of the impossibility of a universal NOT gate for general qubits, most RSP implementations are inherently probabilistic [16,17]; moreover, as the dimension of the remotely prepared state increases, the probability of success becomes smaller. To remotely prepare quantum states deterministically, Alice must instead perform a positive-operator valued measure (POVM) measurement on her quantum state and send the outcome message to Bob [18]. With this larger message from Alice (2 bits for qubit RSP), Bob can transform his state into the state Charles chose using simple unitary operations. Deterministic RSP protocols have been implemented for photon and ion qubit states [19,20,21].

While both QT and RSP allow Alice to communicate quantum information to Bob using shared entanglement and a two-bit classical message, each technique has advantages and disadvantages. Because QT requires a full Bell-state measurement, it is impossible to implement deterministically in linear optical systems [22,23]; in contrast, RSP only requires Alice to make measurements using linear optics (which can be made deterministically). On the other hand, QT does not require even Charles to know what state he is sending to Alice, enabling him to implement entanglement swapping [24], which cannot be accomplished using RSP. For both higher dimensional QT and RSP, the classical communication cost can be shown to scale as $\log_2 d^2$ with the dimension $d$ of the quantum state that has $2d-2$ continuous state-defining parameters (e.g., $\theta$ and $\phi$ in $\cos\theta|0\rangle + e^{i\phi}\sin\theta|1\rangle$) [14]. In this letter we report an implementation of a new quantum communication protocol, known as SuperDense Teleportation (SDT), which has reduced classical information resource requirements compared to QT, simplified measurements for Alice, easier transformations for Bob, and can in principle be implemented deterministically in (linear) optical implementations [25].

## Results

The two state transfer techniques described in the previous section are used to send completely general quantum states. However, it is possible to remotely prepare qubit states that are constrained to lie on a great circle of the Poincaré sphere, requiring only a single bit transferred from Alice to Bob [26,27,28]. Furthermore, this idea of transmitting a state from a constrained portion of Hilbert space may be extended to higher dimensional states [29,30]; the resulting technique, SuperDense teleportation (SDT), can be used to send states at a reduced classical information cost per state parameter [25]. SDT is somewhat similar to the standard RSP protocol in that Charles encodes the state parameters that he wishes to communicate to Bob directly onto Alice's half of the entangled state. However, unlike traditional RSP, instead of attempting to send a general $d$-dimensional state, requiring all $2d-2$ state-defining parameters, Charles only attempts to send a state with $d-1$ state-defining parameters, corresponding to the relative phases of an equimodular state (also known as an equatorial qudit):

$$\left(|0\rangle + e^{i\phi_1}|1\rangle + e^{i\phi_2}|2\rangle + \ldots + e^{i\phi_{d-1}}|d-1\rangle\right)/\sqrt{d}. \tag{1}$$

To do this he applies these phases to the input maximally entangled state, i.e.,

$$\left(|00\rangle + e^{i\phi_1}|11\rangle + e^{i\phi_2}|22\rangle + \ldots + e^{i\phi_{d-1}}|(d-1)(d-1)\rangle\right)/\sqrt{d}, \tag{2}$$

and sends his modified half of the entangled state to Alice. She then measures her qudit ($d$-dimensional quantum state) in a basis which is mutually unbiased to the one Charles used to apply the phases, and sends the measurement outcome, only $\log_2 d$ bits, to Bob, who performs one of $d$ relative-phase-shifting unitary transformations on his particle to recover the intended state (1).

The reduction in classical information required by SDT is not only interesting from a theoretical point of view but is also accompanied by significant experimental simplifications. Chief among these is the reduced complexity of the measurements (e.g., number of interferometers and detectors) that Alice must make on her half of the entangled state. The measurements required for QT are probabilistic when using linear optics; this problem is worsened when teleporting higher dimensional states. Because the percentage of the total higher-dimensional Bell states discriminated with linear optics detection decreases as dimension increases, QT of states $d > 2$ is impossible to do with perfect fidelity without nonlinear interactions [31,32]. Furthermore, although RSP can be performed deterministically for any state dimension, the complexity of the measurement increases quadratically with the state dimension: a $d$-dimensional state with $2d-2$ state parameters requires a POVM with $d^2$ outputs and detectors. SDT, in contrast, requires only a comparatively simple $d$-dimensional mutually unbiased basis measurement to teleport a $d$-dimensional state with $d-1$ state parameters. While SDT sends only half the number of state parameters associated with a $d$-dimensional state, the complexity of the experiment is greatly reduced. For example, the number of detectors scales linearly with the state dimension for SDT instead of quadratically as in RSP. Moreover, the number of different transformations Bob needs to implement are thus also reduced to linear scaling with the dimension—much easier than the quadratic scaling for RSP (and QT). Table 1 summarizes the three protocols.

To experimentally demonstrate SDT's advantages over quantum teleportation and RSP (e.g., reduced classical communication cost and experimental measurement simplification), states with at least two quantum parameters must be transferred [25]. Here we experimentally demonstrate SDT by transmitting equimodular ququart states (4-dimensional quantum states with three independent state parameters). This may be accomplished by preparing entangled states in four modes of one degree of freedom, such as spatial or temporal mode. Instead, however, we use states that are hyperentangled—simultaneously entangled in multiple degrees of freedom—in polarization and orbital angular momentum to produce four-mode entangled states [33].

To create the required hyperentangled states we pump a pair of nonlinear Type-I phase-matched BBO crystals with a 351-nm Ar$^+$ laser (see Fig. 2 and Supplemental Information for details). With rare probability, a high-energy photon may be split by the nonlinear crystals into two lower energy photons through spontaneous parametric downconversion. These crystals were oriented such that a horizontally (vertically) polarized pump photon split in the first (second) crystal will produce two vertically (horizontally) polarized photons. By pumping the crystals with a coherent, equal superposition of horizontal and vertical polarization we created a maximally entangled polarization state [34]. Furthermore, because orbital angular momentum is conserved in the downconversion process, the daughter photons will be correlated in orbital angular momentum as well [35]. By selecting only the $\pm\hbar$ orbital angular momentum modes, we create a state that is maximally entangled in both polarization and spatial mode:

$$\frac{1}{2}(|HH\rangle+|VV\rangle)\otimes(|rl\rangle+|lr\rangle), \quad (3)$$

where $r$ and $l$ are eigenfunctions of the orbital angular momentum operator with $\pm\hbar$ orbital angular momentum [33]. One photon of the resulting state was sent to Charles, and the other to Bob.

To encode the three state parameters that Alice must teleport to Bob, Charles applied phases using liquid crystals and by varying the phase between the two spatial modes, which were processed using a binary forked hologram. These silver-halide holograms were used in conjunction with single-mode fibers to transform the $r$ and $l$ states into two Gaussian modes in the ±1 diffraction orders, respectively (with ~30% efficiency)[36]. After these transformations the total 2-photon entangled state was:

$$\frac{1}{2}(|00\rangle + e^{i\phi_1}|11\rangle + e^{i\phi_2}|22\rangle + e^{i\phi_3}|33\rangle), \quad (4)$$

where $|0\rangle \equiv |Hr\rangle$, $|1\rangle \equiv |Hl\rangle$, $|2\rangle \equiv |Vr\rangle$, and $|3\rangle \equiv |Vl\rangle$ (in reality the labels $r$ and $l$ are reversed for Alice and Bob, but this does not affect the results). Charles then sent the photon on to Alice, who combined the two spatial modes on a polarizing beam splitter to form a "spin-orbit" CNOT gate [37]. By making polarization measurements on the output spatial modes, Alice effectively made measurements in the following basis (which is mutually unbiased to the basis in which Charles applied the phases):

$$|a^\pm\rangle \equiv (|Dr\rangle \pm |Al\rangle)/\sqrt{2}, \quad |b^\pm\rangle \equiv (|Ar\rangle \pm |Dl\rangle)/\sqrt{2}, \quad (5)$$

where $D$ ($A$) is diagonal (anti-diagonal) polarization. The two-photon four-qubit state can then be written as:

$$\frac{1}{4}[|a^+\rangle(-|0\rangle + e^{i\phi_1}|1\rangle + e^{i\phi_2}|2\rangle + e^{i\phi_3}|3\rangle) + |a^-\rangle(|0\rangle - e^{i\phi_1}|1\rangle + e^{i\phi_2}|2\rangle + e^{i\phi_3}|3\rangle) + |b^+\rangle(|0\rangle + e^{i\phi_1}|1\rangle - e^{i\phi_2}|2\rangle + e^{i\phi_3}|3\rangle) + |b^-\rangle(|0\rangle + e^{i\phi_1}|1\rangle + e^{i\phi_2}|2\rangle - e^{i\phi_3}|3\rangle)]; \quad (6)$$

here states refer $|a^\pm\rangle$, $|b^\pm\rangle$ to Alice's photon, while $|0\rangle$, $|1\rangle$, $|2\rangle$, and $|3\rangle$ refer to Bob's. Therefore, Alice's measurement projects Bob's photon into a state that can be corrected by making a π phase shift on the relevant term. In our proof-of-principle experiment, we did not apply these phases for each photon as its partner was detected, which would have required photon storage and feed-forward state correction (see Supplemental Information). Instead, Bob performed a full 2-qubit single-photon tomography on his photon using liquid crystals and a scanning hologram [33]. Bob's hologram, like Charles', was used in conjunction with single-mode fibers to convert a particular spatial mode into a Gaussian mode in the ±1 diffraction orders; using this technique, it was possible to make spatial-mode measurements on the photons in different bases. Since different regions of the hologram (used in conjunction with single-mode fibers) converted different spatial-modes to Gaussians, taking a complete polarization tomography at each hologram region enabled a tomographically over-complete set of polarization and spatial-mode measurements on Bob's photon $(\{H,V,D,A,R,L\} \otimes \{h,v,d,a,r,l\})$, where ($h \equiv (r+l)/\sqrt{2}$, $v \equiv (r-l)/\sqrt{2}$, $d \equiv (r+il)/\sqrt{2}$, and $a \equiv (r-il)/\sqrt{2}$) [33]. Correlating Bob's measurement outcomes with Alice's, we used maximum likelihood state reconstruction [38] to determine what state $\rho$ Bob received for each of Alice's measurement outcomes. Finally, we then numerically applied the transformation indicated by Alice's measurement outcome to the reconstructed states, to compare with the original state intended to be transmitted.

The average fidelity over all the measured teleported states (see Fig. 3 and 4) was 87.0(1)%, approximately twice the 44% average fidelity limit for sending a single equimodular ququart state over a classical channel without entanglement (see Methods). For comparison, perfect QT of a qubit exceeds the classical limit by $\Delta F_{qubit} \equiv F_{quantum}^{average} - F_{classical}^{average} = 1 - 2/3 = 1/3$ [40], and actual achieved results are lower, often much lower. In addition, recent improvements in spatial-mode sorting could increase the fidelity of SDT even further [39].

As seen in figure 5, the diagonal elements of our reconstructed density matrices are not all equal, in contrast to the theoretical expectation for equimodular states (see Methods). This inequality appears to arise from spatial-mode crosstalk in both Alice and Bob's measurements; such crosstalk is the main limitation in the fidelity of the reconstructed states. We also examined how well each of the phases that Charles sent was transferred from Alice to Bob. From our state reconstructions, we estimate that the systematic error in the phases of Bob's reconstructed states was ±4.0° for each $\phi_a$, $\phi_b$, and $\phi_c$ (see Fig. 2 for definitions of these phases in terms of $\phi_1$, $\phi_2$, and $\phi_3$). This deviation suggests Charles and Alice can reliably communicate nearly $10^5 \left(= \left(\frac{360}{2*4.0}\right)^3\right)$ distinguishable states to Bob.

In addition to the full state tomographies, we also made partial reconstructions over a much larger number of input phases, in order to verify that Charles and Alice could teleport a wide range of phase settings to Bob. For these measurements, Charles varied one of the three phases while keeping the others constant. Then, instead of making all 36 measurement configurations for a full two-qubit tomography for each of Charles's phase settings, Bob only made specific measurements to find the values of the three interferometric functions which varied with phase ($\langle Hh|\rho|Hh\rangle$, $\langle Dr|\rho|Dr\rangle$, and $\langle Dl|\rho|Dl\rangle$, where $\rho$ is the state of Bob's photon). Each of these measurements varied uniquely with each of the phases Charles applied, resulting in phase-dependent fringe curves (see Fig. 6). Some measurements displayed unexpected phase dependence, varying with phases of which they were supposedly theoretically independent. This deviation from the expected measurement/phase relationship is further evidence that spatial-mode crosstalk is a limiting factor in this proof-of-principle experiment.

**Discussion**

All improvements in both classical communication cost and experimental simplification can be associated with the shape of the constrained space in which the equimodular states reside, specifically, a type of hyper-torus, which is topologically different from the space associated with general quantum states of the same number of parameters. Because of this topological difference, it is possible to perform universal (within the restricted portion of the space) NOT gates, which are impossible to implement for general quantum states [15]. As was previously mentioned for qubit RSP, it is the impossibility of this operation over general quantum states that requires Alice to use a POVM and two classical bits in RSP in order to send Bob enough information to transform his state to the target state. In one-parameter SDT, the "universal" NOT gate (mapping all one-dimensional equimodular states to their orthogonal) required for Bob to recover the target state is just a simple π-phase shift between two basis states. When moving to higher dimensional spaces, Bob must be able to perform an entire set of universal NOT gates that transform an input state to each of its orthogonal states. Again, these transformations are impossible to implement for general qudit states, but are simple relative phase shifts for inputs restricted to the set equimodular states used in SDT.

The topological structure of equimodular states also influences their information content. In particular, the parameters of an equimodular state sweep out a more significant portion of Hilbert space than an equivalent number of parameters in a lower dimensional general quantum state (e.g., $\frac{4\pi^2}{3}$ versus $\frac{3\sqrt{\pi}}{4}$ for two-parameter state communication using SDT or QT, respectively)(see Methods and Supplemental Information). In fact, the ratio of the volume of the space of equimodular states to the corresponding space of general quantum states grows exponentially with the number of state parameters (see Methods). This volume ratio is related to the ratio of the number of states that can be "packed" into the two volumes: as the dimension increases, an exponentially larger number of statistically distinguishable states (for a small minimum statistical distance between states) can be packed into the class of equimodular states (hyper-torus) than into the class of general states (hyper-sphere) with the same number of state parameters. A second perspective of why equimodular states have greater information content can be understood by examining the amount of information that can be inferred about general and equimodular state from a single-shot measurement. We define *classical teleportation* as the optimal strategy for guessing a quantum state given a single-shot measurement [40], which is equivalent to the optimal strategy for Alice to communicate an unknown state to Bob by sending the result of a single measurement without shared entanglement. With this definition, the average fidelity of classical teleportation is lower for the equimodular states used in SDT $\left(\frac{2N+1}{(N+1)^2}\right)$ than for general quantum states $\left(\frac{4}{N+4}\right)$ of the same number of quantum parameters (*N*) used in quantum teleportation and RSP (see Methods) [41]. This is an indication that SDT not only requires transmitting fewer classical bits to teleport the quantum parameters, but that these parameters in some sense contain more information on average than the parameters of general states used in RSP and QT.

We have implemented a novel entanglement-enhanced quantum state communication protocol which can communicate quantum state parameters with less classical information transfer and simpler measurements than standard quantum teleportation or RSP. Using SDT we were able to transfer a wide variety of states from Alice to Bob with much better fidelity than classical teleportation. In addition to the pure target states that were teleported in this experiment, these techniques might also be extended to transfer partially mixed equimodular states as well. We also speculate that SDT might be used to exchange quantum state inputs between a client and quantum server in blind quantum computing [42]. It should be noted that both SDT and RSP are closely related to quantum steering [43]. We are currently investigating the application of recent advances in quantum steering and semidefinite programming to the quantum states reconstructed in this experiment [44,45]. Because universal NOT operations can be performed on equimodular states, they might also have interesting applications in ideal quantum cloning [46] and in dynamical decoupling noise-reduction techniques [47]. Finally, this research shows that equimodular states have topological features that might make them superior to general states for quantum state communication (i.e., the power of SDT comes from the fact that equimodular states are topologically different from general quantum states) and motivates further investigations into how such constrained states might be used to optimize other quantum information techniques. For example, equimodular states are precisely those necessary to implement the quantum "fingerprinting" [48].

## Methods
### I. State Reconstruction
A full two-qubit polarization and spatial-mode tomographic reconstruction was performed for each state that was transmitted from Alice to Bob using superdense teleportation. The density matrices representing these states were then calculated using maximum likelihood state estimation techniques [38]. However, each of Alice's detectors heralds a different state on Bob's side, so a simple reconstruction on Bob's photon without accounting for Alice's measurement yields a mixed state. To reconstruct the state that Bob would have measured had he made the corrective unitary transformation on his photon based on Alice's message, state tomographies were performed in coincidence with Alice's measurements (see Fig. 5; the four matrices (one for each of Alice's measurements) were then averaged after numerically applying the respective unitary transformations (see Fig. 4). The phase angles (corresponding to the parameters that Charles encoded) and the fidelity with the target state were calculated for each of the resulting density matrices (see Table 2).

## II. Packing Number and Volume Ratio

One way to measure the complexity of a set of states is to consider the "packing number" with respect to the Bures distance [49]. For states with an angle less than $\pi/2$, the Bures distance coincides with the usual Euclidean distance, which will be used here for simplicity.

Given a subset $T$ in a $d$-dimensional real Euclidean space $\mathbb{R}^d$, we define the *packing number* $P(T,\delta) = \max N$ as the maximal number of points $x_1,...,x_N$ in $T$ such that $\|x_j - x_k\| > \delta$ for all $1 \leq j \neq k \leq N$. If $T$ is sufficiently smooth and of dimension $n$, we have:

$$\lim_{\delta \to 0} \left(\frac{\delta}{2}\right)^n P(T,\delta) = c(n)\mathrm{vol}_n(T). \tag{7}$$

Here $\mathrm{vol}_n(T)$ is the $n$-dimensional measure [50] and $c(n)$ is the packing density of Euclidean space. It is known [51] that $\frac{\zeta(n)}{2^n} \leq c(n) \leq 1$ holds for the Riemann $\zeta$-function. The estimate $c(1) = 1$ is easy, and $c(2) = \frac{\pi}{\sqrt{18}}$ is due to Gauss. For larger dimension, only lower and upper estimates are known, see e.g., [52].

For the set of equimodular states $T_n = n^{-1/2}\mathbb{T}^n \subset \mathbb{C}^n$, $d = n-1$ (because $\mathbb{C}^n = \mathbb{R}^{2n}$ and there are $n-1$ real parameters in $T_n$), we find

$$\mathrm{vol}_d(T_n) = n^{-\frac{n-1}{2}}(2\pi)^{n-1}, \tag{8}$$

where we divide by $2\pi$ to account for the fact that a global phase does not change the state. Let us compare this with a sphere of dimension $d$, i.e., the set of vectors $S^d \subset \mathbb{R}^{d+1}$ of points of length 1. Let us denote by $B_{d+1}$ the unit ball in $\mathbb{R}^{d+1}$, i.e., the set of all points of length less than one. It is well-known that

$$\mathrm{vol}_{d+1}(B_{d+1}) = \frac{\pi^{(d+1)/2}}{\Gamma\left(\frac{d+1}{2}+1\right)}. \tag{9}$$

Using polar coordinates and Stirlings formula $\left(\Gamma(z+1) = z! \sim \sqrt{2\pi z}\left(\frac{z}{e}\right)^z\right)$ this implies

$$\mathrm{vol}_d(S^d) = \frac{(d+1)\pi^{\frac{d+1}{2}}}{\Gamma\left(\frac{d+1}{2}+1\right)} \sim \frac{(d+1)\pi^{\frac{d+1}{2}}}{\sqrt{\pi(d+1)}}\left(\frac{2e}{d+1}\right)^{\frac{d+1}{2}}$$
$$\sim \sqrt{2}(2\pi e)^{d/2} d^{-d/2} \tag{10}$$

Here the $\sim$ symbol denotes that if $a_d \sim b_d$ then $\lim_{d \to \infty} a_d/b_d = 1$. The error of this approximation can be reduced by including higher-order terms when approximating the $\Gamma$-function. For $d = n-1$ we find

$$\text{vol}_d(T_n) = (2\pi)^{n-1} n^{-\frac{n-1}{2}}$$
$$> \sqrt{2}(2\pi e)^{\frac{n-1}{2}} (n-1)^{-\frac{n-1}{2}} \sim \sqrt{2e}\left(\sqrt{2\pi e}\right)^{n-1} n^{-\frac{n-1}{2}} = \text{vol}_d(S^d),$$
(11)

because $2\pi > \sqrt{2\pi e}$. Thus, the packing number for a torus is larger than the corresponding packing number for a sphere of the same dimension, assuming small $\delta$. For these calculations, we have assumed that the general class of states is embedded in a sphere instead of a complex projective space. However, a full calculation in complex projective space still shows that the ratio of the volumes of equimodular versus general states of the same number of parameters grows exponentially (See Supplemental Information). In realistic quantum communication experiments, systematic error and a limited number of state copies will constrain the minimum statistical distance ($\delta$) which two states can be separated and still be experimentally distinguished. To address this issue, we have also considered the packing number for equimodular versus general quantum states for a fixed threshold $\delta$ (See Supplemental Information).

### III. Classical Teleportation Fidelity

To establish the optimal average fidelity with which a *d*-dimensional quantum state can be transmitted over a classical channel without entanglement, we must determine how well the state can be estimated with a single optimal measurement. Alice then makes this measurement and sends the result to Bob (who knows Alice's measurement strategy), who makes a state estimation based on this message. The best average fidelity that one can achieve using this classical teleportation strategy has been calculated to be [41]:

$$F_{general} = \frac{2}{1+d} = \frac{4}{4+N} \sim \frac{4}{N}.$$
(12)

Because the hyper area of an equimodular state is larger than that for a general quantum state of the same number of parameters, the former are more difficult to send classically. The average fidelity of an optimal state estimation strategy can be calculated using well-established methods of Massar and Popescu [40]. It is optimal to measure in a basis which is mutually unbiased to the basis in which Charles applies the phases, for example:

$$|k\rangle \equiv \frac{1}{\sqrt{d}} \sum_{j=0}^{d-1} e^{i\frac{2\pi jk}{d}} |j\rangle.$$
(13)

If Alice measures in this basis and sends the result to Bob, it is the optimal strategy for Bob to simply guess the same state that Alice measured. The average fidelity of his state is then:

$$F_{unimodular} = \sum_{k=0}^{d-1} \int \frac{d\vec{\phi}}{(2\pi)^d} F(k,\vec{\phi}) P(k,\vec{\phi}) = \frac{1}{d^4} \sum_{k=0}^{d-1} \int \frac{d\vec{\phi}}{(2\pi)^d} \left| \sum_{j=0}^{d-1} e^{i\left(\phi_j - \frac{2\pi jk}{d}\right)} \right|^4,$$
(14)

where $P(k,\vec{\phi})$ is the probability that Alice detects state *k* of the measurement basis and $F(k,\vec{\phi})$ is the fidelity of Bob's guess, given that Alice measured state *k*. After shifting phase angles by $\frac{2\pi jk}{d}$ (by a simple redefinition), we simplify the fidelity to:

$$F_{unimodular} = \frac{1}{d^3} \int \frac{d\vec{\phi}}{(2\pi)^d} \left| \sum_{j=0}^{d-1} e^{i\phi_j} \right|^4 = \frac{1}{d^3} \int \frac{d\vec{\phi}}{(2\pi)^d} \sum_{i,j,k,l} e^{i(\phi_i + \phi_j - \phi_k - \phi_l)}. \tag{15}$$

The integral is unity for all phase combinations that add to zero, and zero for all other combinations:

$$F_{unimodular} = \frac{1}{d^3} \sum_{i+j-k-l=0} 1. \tag{16}$$

These indexes add to zero when *i=k* and *j=l* (occurs $d^2$ times) and when *i=l* and *j=k* (occurs $d^2$ times); however, this double-counts when *i=k=j=l* (occurs $d$ times). Therefore, the optimal average fidelity for an equimodular state becomes:

$$F_{unimodular} = \frac{2d^2 - d}{d^3} = \frac{2N+1}{(N+1)^2}. \tag{17}$$

Thus, we see that the average classical teleportation fidelity for equimodular ququart states is 44%. Examining the asymptotic behavior of the average fidelity for large *N* ($F_{unimodular} \sim \frac{2}{N}$) of general states (eqn. 12) versus equimodular states (eqn. 17), we see that equimodular states can be transmitted over a classical channel (without entanglement) with on average half the fidelity that general states (with the same number of state parameters) can be transmitted.

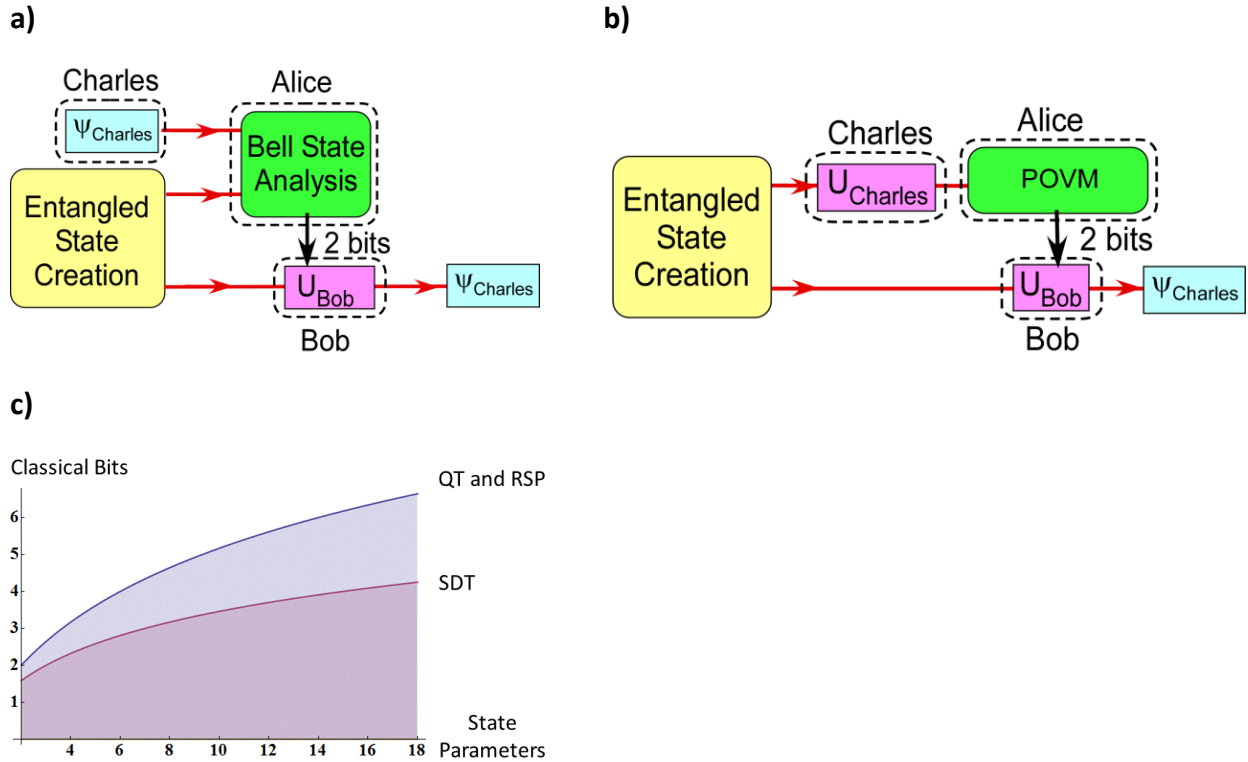

**Figure 1 | Schemes to transfer one qubit. a)** Quantum teleportation layout. Charles prepares a state for Alice, who performs a Bell measurement between her state and Charles'. She then transmits the outcome to Bob, who is able to transform his photon into the state Charles had chosen. **b)** RSP layout. Charles performs a unitary transformation on one photon and sends it to Alice, who makes a POVM measurement on the state. She then sends the outcome to Bob, who transforms his state into the state Charles chose. **c)** The required number of transmitted classical bits for QT, RSP and SDT as a function of the number of parameters teleported; for a large number of parameters, the ratio of classical bits needed for QT and RSP to bits needed for SDT approaches 2.

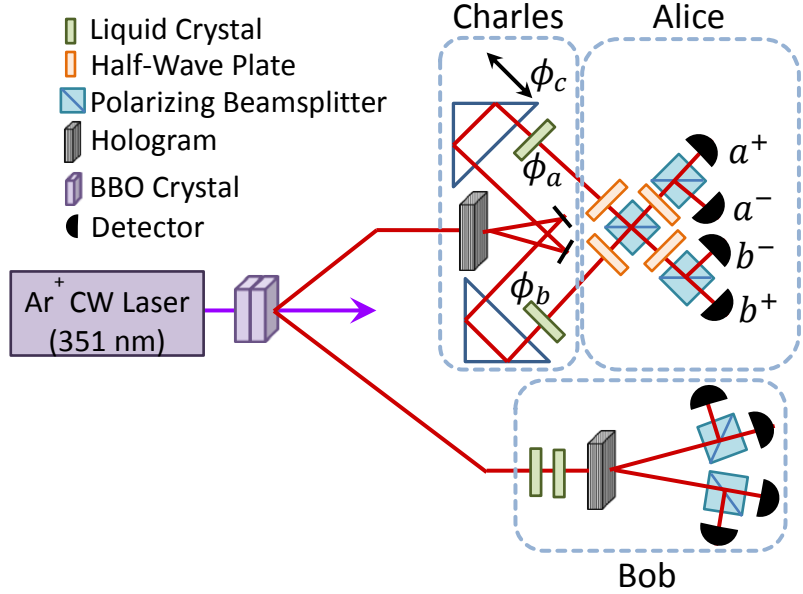

**Figure 2 | Experimental setup for our SDT implementation.** Charles applies phases using liquid crystals and adjusting an interferometer path length. These phases are linear combinations of the phases given in Eq. 1 ($\phi_a = \phi_1 - \phi_2 + \pi/2$, $\phi_b = \phi_2 - \pi/2$, and $\phi_c = \phi_3 + \pi/2$), but still span the space of equimodular states that Charles can prepare. Alice then makes a single-photon two-qubit Bell-state measurement on the polarization and spatial-mode of her photon [37]. By measuring in coincidence with Alice, Bob can determine the state heralded by each of Alice's measurements.

**Table 1 | Resources required to send N state parameters with 100% fidelity for each technique using linear optics**

|  | State dimension | Success probability | Classical bits | Alice Detector # | Bob Transformation # | Known to Charles |
|---|---|---|---|---|---|---|
| QT | 2* | 1/2 | 2 | 4 | 4 | optional |
| RSP (probabilistic) | (N+2)/2 | 2/(N+2) | 1 | 1 | 1 | required |
| RSP (deterministic) | (N+2)/2 | 1 | $2\log_2[(N+2)/2]$ | $[(N+2)/2]^2$ | $[(N+2)/2]^2$ | required |
| SDT | N+1 | 1 | $\log_2[N+1]$ | N+1 | N+1 | required |

*QT of dimension $d > 2$ requires nonlinear optics, or addition of $d$ entangled ancilla qubits [53].

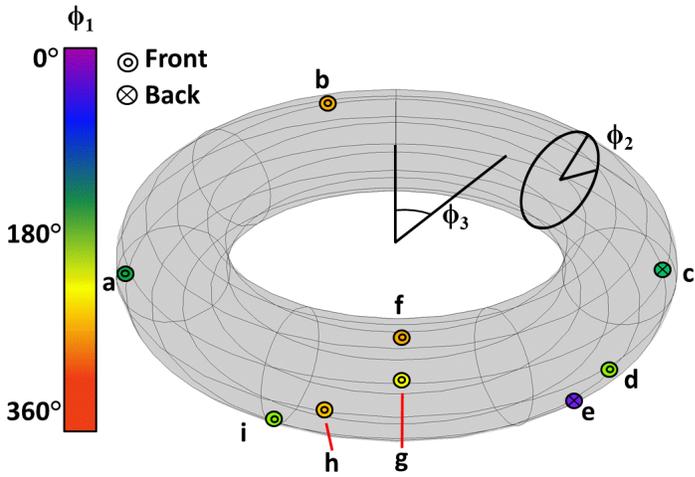

**Figure 3 | A visual representation of the distributions of the states we communicated from Alice to Bob using SDT.** The states that were measured can be represented as lying on a three-dimensional hyper-torus (one dimension for each state parameter) embedded in a six-dimensional Euclidean space for general ququart states. The average fidelity of all teleported states was 87.0(1)%. "Front" in the legend refers to point locations on the side towards the viewer in this perspective while "Back" refers to those obscured by the front surface. The $\phi_1$ parameter can be read from the fill-color of the circle surrounding the point locator on the surface of the torus parameterized by $\phi_2$ and $\phi_3$.

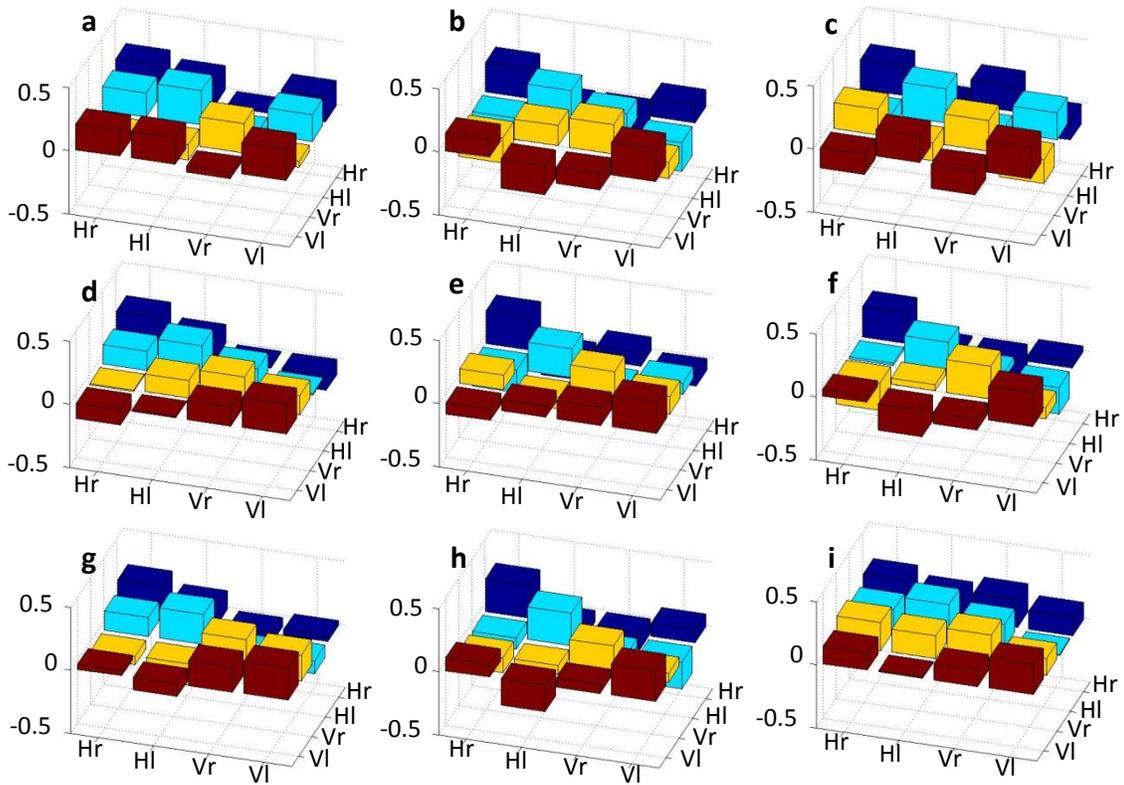

**Figure 4 | Reconstructed density matrices of transmitted states.** The real parts of the reconstructed density matrices are displayed for each state transmitted from Alice to Bob with superdense teleportation (after numerical correction). The letter label on each sub-figure corresponds to the same letter in Figure 3 and Table 2.

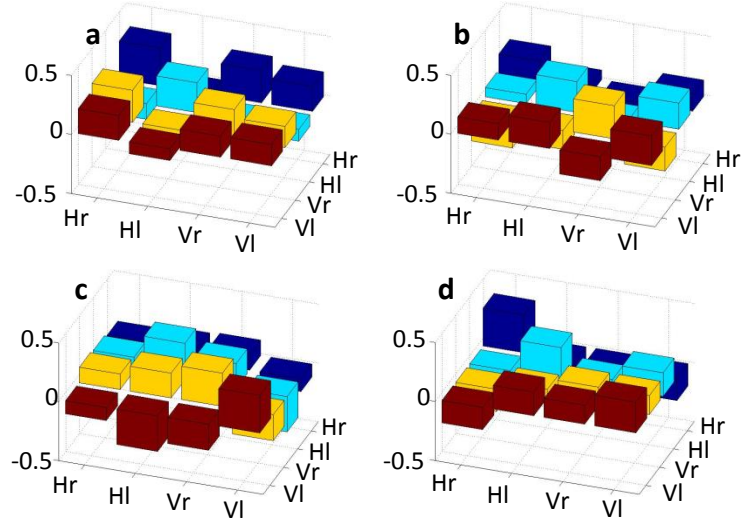

**Figure 5 | Reconstructed density matrices without Bob's corrections.** This is an example of the reconstructed density matrices (real parts) that Bob receives when measured in coincidence with each of Alice's respective measurement outcomes. The inequality of the diagonal elements indicates cross-talk in Alice and Bob's measurements, the major limiting factor of the fidelity in this experiment.

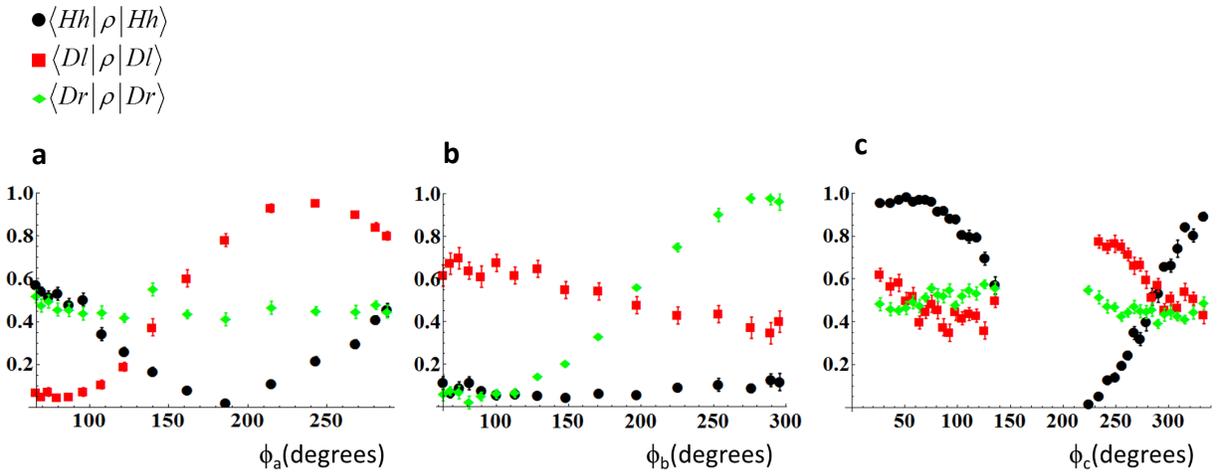

**Figure 6 | Measurement fringes as a function of Charles' phases.** For each figure only one phase is varied while the others are fixed. In figure **a)** the theoretical fidelity range ($\Delta f \equiv$ maximum-minimum) for the black, red, and green curves are $\Delta f = 0.5$, 1, and 0, respectively; in **b)** the theoretical fidelity range values are $\Delta f = 0$, 0, and 1; and in **c)** theoretical fidelity range values are $\Delta f = 1$, 0, and 0. The larger-than-predicted fidelity range of some of these curves arises from cross-talk in the polarization/spatial-mode measurements. The missing strip of data around 180° in the final figure was due to instability in the active feedback system used to stabilize our interferometer at angles near this value of $\phi_c$.

## Table 2 | Summary of experimental results

| Figure | Target Phases (°) | Measured Phases (°) | Ave. Fidelity (%) with target state |
|---|---|---|---|
| a | 112, 180, 278 | 109.7(5), 176.0(6), 283.7(5) | 86.2(3) |
| b | 270, 90, 324 | 266.4(6), 80.7(7), 309.5(7) | 85.7(3) |
| c | 112, 277, 119 | 113.0(5), 272.4(6), 122.9(6) | 87.8(3) |
| d | 180, 180, 137 | 175.6(5), 176.8(5), 141.2(6) | 86.9(3) |
| e | 26, 202, 145 | 23.7(4), 204.4(5), 154.9(4) | 86.4(2) |
| f | 270, 90, 184 | 262.0(7), 80.8(6), 193.7(7) | 86.8(3) |
| g | 211, 158, 185 | 208.5(4), 162.9(4), 191.0(4) | 88.4(3) |
| h | 268, 148, 209 | 273.2(5), 141.4(5), 208.9(6) | 86.2(3) |
| i | 180, 277, 223 | 176.4(5), 272.6(5), 222.6(6) | 89.2(2) |

# Supplemental Information for SuperDense Teleportation using Hyperentangled Photons

**Trent M. Graham[1], Herbert J. Bernstein[2], Tzu-Chieh Wei[3], Marius Junge[4], Paul G. Kwiat[1]**

[1]*Department of Physics, University of Illinois at Urbana-Champaign, 1110 W. Green Street, Urbana, IL 61801, USA*
[2]*School of Natural Science and ISIS Institute for Science and Interdisciplinary Studies, Hampshire College, Amherst, MA 01002, USA*
[3]*C.N. Yang Institute for Theoretical Physics and Department of Physics and Astronomy, State University of New York at Stony Brook, Stony Brook, NY, USA*
[4]*Department of Mathematics, University of Illinois at Urbana-Champaign, 1409 W. Green Street, Urbana, IL 61801, USA*
*Author e-mail address: tgraham2@illinois.edu*

## I. Source Details

The source of entangled photons used in our experiment was created by pumping two orthogonally oriented 0.6-mm-thick BBO crystals with a 351-nm Ar$^+$ laser focused to a 90-μm beam waist. The optic axis of the BBO crystals was oriented such that the 702-nm downconversion photons exited the crystal with a half-opening angle of 3°. The downconversion was then filtered by 10-nm FWHM interference filters and measured using PerkinElmer single photon counting modules with a coincidence window of 10 ns. For states b, f, g, and e (as referenced in Fig. 3 in the main text) hyperentangled photons were generated and measured with a total coincidence rate of ~90 s$^{-1}$ into all detected single modes, with a ~0.6% heralding efficiency of the teleported state. For states a, c, d, h, and i, a different pump power was used resulting in measured coincidence rate of ~140 s$^{-1}$ and a heralding efficiency of ~0.5%. The majority of loss in the system arises from transmission through holograms and liquid crystals, imperfect coupling into single-mode-fibers, and imperfect detection by the single photon counting modules.

## II. Feed-Forward Corrections

While we did not perform feed-forward state correction in this experiment, we can outline one possible way to extend our implementation to allow Bob to perform corrective transformations on his photons based on Alice's measurement outcomes. For this extension, two main modifications must be made to our implementation (see Supplementary Fig. 1). First, Bob must have a way of storing his photon to allow Alice time to measure her photon and transmit the outcome to Bob. Perhaps the easiest way to implement such a delay is to use an optical delay line. Secondly, Bob must possess a way of quick applying the corrective unitary transformations in the path of his photon allowing him to convert his photon to the target state based on Alice's message. Bob could make these transformations quickly using Pockels cells placed before and after his hologram and thus perform full feed-forward correction of his state.

## III. Volume Estimates

The estimates in the Methods section used the simplified assumption that the set of states is embedded in a sphere. More correctly, we should perform the same calculations for the projective space $CP_m = S^{2m-1}/\mathbb{T}$ of rank-one density, where $CP_m$ denotes an $m$-dimensional complex projective space (corresponding to an $m$-dimensional general quantum state), $S^{2m-1}$ represents a (*2m-1*)-dimensional sphere, and $\mathbb{T}$ is the compact unitary group (also known as $U(1)$). In this equation, we identify vectors with the same global phase factor. We now show that using projective spaces leads to similar estimates as calculated in the manuscript. A key point is that if a space can be identified as the quotient of unimodular groups, then via the Haar measure (the respective groups' invariant distance measure), its volume is the ratio of the volumes of the corresponding groups.

The volume calculation for equimodular states can be readily computed from the definition of these states. Equimodular states of dimension $n$ are defined by points on $T_n = n^{-1/2} \mathbb{T}^n \subset \mathbb{C}^n$, up to a global phase factor, where points in $\mathbb{T}^n$ are labeled by $(e^{i\phi_1}, e^{i\phi_2}, \cdots, e^{i\phi_n})$. Integrating over these phases and dividing by $T_1$ to account for the fact that an overall phase factor does not change the state, we see that the volume of equimodular states is given by

$$\mathrm{vol}_{n-1}(T_n / T_1) = (2\pi)^{n-1} n^{-(n-1)/2}, \tag{1}$$

where the $n-1$ subscript refers to the dimension of the space that the volume is calculated in.

To compute the volume of general states, we first need to find the Haar measure on the surface of a $2m$-dimensional sphere (i.e., $S^{2m-1} \subset \mathbb{R}^{2m}$). This can be obtained from identifying $S^{2m-1}$ as a quotient of two orthogonal groups $O_{2m}/O_{2m-1}$. As multiplication by a phase factor in $S^{2m-1}$ is again a group action of the compact group $\mathbb{T}$, the complex projective space can be identified as a quotient of groups $CP_m = O_{2m}/\mathbb{T} \times O_{2m-1}$. Therefore [54, 55], we deduce that the volume of the set of $m$-dimensional general states is

$$\begin{aligned}
\mathrm{vol}_{2m-2}(CP_m) &= \frac{\mathrm{vol}_{2m-1}(S^{2m-1})}{2\pi} = \frac{(2m-1)\dfrac{\pi^{(2m-1)/2}}{\Gamma\left(\dfrac{2m-1+1}{2}+1\right)}}{2\pi} \\
&= \frac{(2m-1)}{2}\frac{\pi^{(2m-3)/2}}{m!} \\
&\sim \frac{(2m-1)\pi^{(2m-3)/2}}{2\sqrt{2\pi m}} e^m m^{-m} \\
&\sim \frac{1}{\sqrt{2\pi^2}} m^{-1/2} (\pi e)^m m^{-m},
\end{aligned} \tag{2}$$

where $\Gamma$ denotes the gamma function and Stirling's approximation was used to simplify the factorial.

We can now compare volumes for equimodular (Eqn. 1) and general states (Eqn. 2) with the same number of state parameters $N \equiv 2(m-1) = n-1$. Thus, for $n$ odd we use $m = 1 + \dfrac{n-1}{2}$ and find the volume for general states is

$$\mathrm{vol}_{2m-2}(CP_m) \sim \frac{\pi e}{\sqrt{2\pi^2}}\left(1+\frac{n-1}{2}\right)^{-1/2}(\pi e)^{(n-1)/2}\left(1+\frac{n-1}{2}\right)^{-\left(1+\frac{n-1}{2}\right)}$$
$$\sim \frac{\pi e}{\sqrt{2\pi^2}}\left(\frac{n+1}{2}\right)^{-3/2}(2\pi e)^{(n-1)/2}n^{-(n-1)/2}.$$
(3)

Comparing the volume of the set of equimodular states (Eqn. 1) to the volume of the set of general states with the same number of state parameters (Eqn. 3), we find the ratio of the two volumes is

$$\frac{\mathrm{vol}_{n-1}(T_n/T_1)}{\mathrm{vol}_{n-1}\left(CP_{1+\frac{n-1}{2}}\right)} = \frac{(2\pi)^{n-1}n^{-(n-1)/2}}{\frac{\pi e}{\sqrt{2\pi^2}}\left(\frac{n+1}{2}\right)^{-3/2}(2\pi e)^{(n-1)/2}n^{-(n-1)/2}}$$
$$\sim \frac{(2\pi)^{n-1}}{(\sqrt{2\pi e})^{n-1}}$$
(4)

where again we observe that $2\pi > \sqrt{2\pi e}$ implies that the volume of the equimodular states in projective space is larger than the volume of general state space with the same number of parameters. Let us note in passing that the volumes for the sphere and the projective space have the same leading term for large $n$.

As a conclusion, we see that, confirming the calculation in the manuscript, the equimodular states occupy a larger volume, and hence more of these states can be statistically distinguished for small $\delta$, the minimum distance between the packed states. Some of this advantage, however, is obscured by the difficulty of finding the optimal packing configuration and the unknown density of this packing configuration. Indeed, the volumes differ by at most a factor $2^n$, and hence, in practice, we would need very precise configurations to realize this advantage. Finding optimal packing configurations is a mathematically hard problem which remains unsolved despite centuries of research.

**IV. General Packing Numbers and Distinguishability of Equimodular versus General States**

The preceding volume approach only determines packing numbers for infinitesimal values of $\delta$, and the rate of convergence may depend heavily on the *shape* of the manifold. Experimentally, limited numbers of state copies as well as systematic noise will increase $\delta$. We will now show that for an equal number of state parameters, there are $2^{c_1 n}$ general states and $2^{c_2 n}$ equimodular states separated by a minimum distance $\delta$, where $c_1$, $c_2$, and $\delta$ are constants which are independent of the state dimension. For technical reasons it is better to work with the so-called entropy numbers, defined as follows: for some set $K \subset \mathbb{R}^d$ the *entropy number* of that set is $N(K,\varepsilon) = \min k$, where there are points $x_1, ..., x_k \in \mathbb{R}^d$ with

$$K \subset \bigcup_i x_i + \varepsilon B_d.$$
(5)

Here $B_d = \{x\,|\,\|x\| \leq 1\} \subset \mathbb{R}^d$ is the $d$-dimensional unit ball. Intuitively, entropy numbers are the minimal number of overlapping $\varepsilon$-radius balls to cover the set K. In contrast, the packing number, $P(K,\varepsilon)$, is intuitively the *maximal* number of $\varepsilon/2$-radius balls (thus the distance between any two balls is at least $\varepsilon$) that can be squeezed inside K. It is thus easy to see that $P(K,2\varepsilon) \leq N(K,\varepsilon)$. The "pigeonhole principle" [56] implies $N(K,\varepsilon) \leq P(K,\varepsilon)$. Thus, it suffices to study entropy numbers to estimate the scaling of packing numbers. It turns out that a lower bound on the entropy numbers can be obtained via the so-called Sudakov's inequality [57], which when applied to $T_n = n^{-1/2}\mathbb{T}^n \subset \mathbb{C}^n = \mathbb{R}^{2n}$, yields

$$\sqrt{\frac{2n}{\pi}} \leq C\int_0^{diam(K)} \sqrt{\log_2 N(K,\varepsilon)}d\varepsilon, \tag{6}$$

for some unknown constant $C$ which depends on the geometry of the complex projective space.

We will now estimate the integral in eqn. (6) using the following volume estimate [57]

$$N(B_n,\varepsilon) \leq \left(1+\frac{2}{\varepsilon}\right)^n. \tag{7}$$

Using logarithmic identities and setting $K \subset B_d$ we have

$$\sqrt{\log_2 N(K,\varepsilon)} \leq \sqrt{\frac{2d}{\ln 2}}\sqrt{\frac{1}{\varepsilon}}. \tag{8}$$

To match the form of Sudakov's inequality in equation 6, we integrate equation 8 over epsilon. Breaking the integral into two intervals of [0, $\delta$] and [$\delta$, $D$] (where $D$ is the diameter of set $K$) and setting $d = 2n$ and $D \leq 2$, we have

$$C\int_0^{diam(K)} \sqrt{\log N(K,\varepsilon)}d\varepsilon = C\int_0^{\delta} \sqrt{\log N(K,\varepsilon)}d\varepsilon + C\int_\delta^D \sqrt{\log N(K,\varepsilon)}d\varepsilon$$

$$\leq \sqrt{\frac{4n}{\ln 2}}C\int_0^\delta \sqrt{\frac{1}{\varepsilon}}d\varepsilon + (2-\delta)C\sqrt{\log N(K,\delta)} \tag{9}$$

$$\leq C\sqrt{\frac{4n}{\ln 2}}2\delta^{1/2} + 2C\sqrt{\log_2 N(K,\delta)}.$$

Thus, we have

$$\sqrt{\frac{2n}{\pi}} \leq C\sqrt{\frac{4n}{\ln 2}}2\delta^{1/2} + 2C\sqrt{\log_2 N(K,\delta)}. \tag{10}$$

It is now possible to find a $\delta$ independent of dimension $n$ such that the inequality in equation (10) holds. For example, taking the first term on the right-hand side to be equal or smaller than half the part of the expression on the left-hand side, $C\sqrt{\frac{4n}{\ln 2}}2\delta^{1/2} \leq \frac{1}{2}\sqrt{\frac{2n}{\pi}}$, i.e. $\delta \leq \frac{\ln 2}{32\pi C^2}$, we find $\sqrt{\frac{n}{8\pi C^2}} \leq \sqrt{\log_2 N(K,\delta)}$. This, combined with equation (7) and the fact that $P(K,2\delta) \leq N(K,\delta)$, leads to the following result:

**Theorem:** There exists a $\delta$ and $C > 0$ (independent of $n$) such that the packing number for general quantum states is bounded

$$e^{\frac{2n}{\delta}} \geq P(K, \delta) \geq 2^{\frac{n}{8\pi C^2}}. \tag{11}$$

This means that if we can statistically distinguish states separated by a distance $\delta$, than we can encode at least $\frac{n}{8\pi C^2} \equiv c_1 n$ many bits for $\delta \leq \frac{\ln 2}{32\pi C^2}$. Using similar arguments one can show that even considering restricted equimodular states with phases restricted to 0 or π, it is still possible to encode bits.

In summary, using the freedom of encoding in the whole sphere we see that only $c_1(\delta)n$ many bits can be encoded with mutual distance $\delta$. Despite not knowing the exact behavior of packing numbers, we can determine that for some minimal statistical distance between states, there is a constant ratio of bits that can be encoded in each volume, which is independent of dimension.

## V. Measurement Outcomes of Equimodular States

In the previous two sections, we examined the number of states that could be packed into the class of equimodular states compared with general quantum states with the same number of state parameters (but different state dimension). It is also instructive to compare equimodular states to general states with the *same* state dimension (and therefore double the number of state parameters). Unfortunately, these two classes of states are represented by shapes with different Euclidean dimension, since they have a different number of state parameters, making a simple volume comparison of the two shapes not very informative. A potentially more instructive comparison can be made by examining how measurements on states are affected if a state is constrained to be equimodular. The probability that a measurement on a state yields a particular result reveals information about that state. By examining how the outcome probabilities of different measurements are constrained for equimodular states, we can directly compare equimodular and general states. For example, if measurements are made in the same basis in which Charles applied his phases, then all equimodular states will give the same measurement probability signature, i.e., uniform count probability for all outcomes. This is much more constrained than general states, which can have an arbitrary normalized measurement outcome probability. However, measurements made on an equimodular state in a mutually unbiased basis will vary with the relative phase, though these measurements are still constrained as compared to general states, and the severity of these constraints will depend on the dimension of the equimodular state. For example, just like a general qubit, measurements on an equimodular qubit state can have completely arbitrary normalized outcome probabilities. However, the ratio of possible measurement outcome probabilities of equimodular states compared to general states with the same dimension decreases as dimension increases. As an example, an equimodular qutrit (2 free parameters) accesses only ~31% of the measurement outcome probability combinations of general qutrit states (4 free parameters) (see Supplementary Fig. 2). Calculating the region of outcome probabilities accessible to equimodular states becomes more difficult as the dimension of the state increases; the general calculation is beyond the scope of this work.

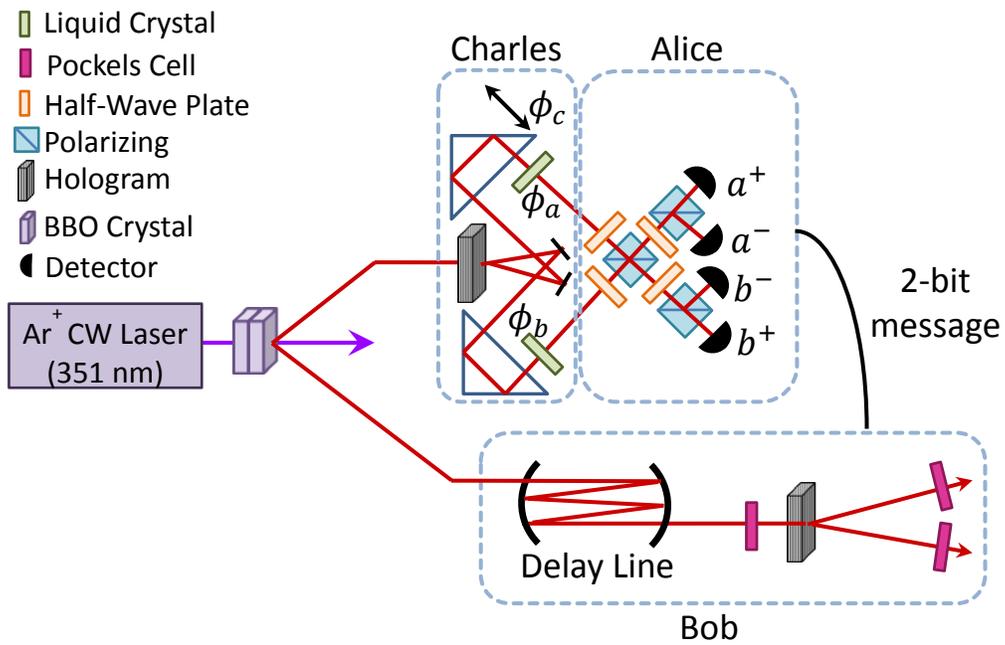

**Supplementary Figure 1 | A possible layout for implementing SDT with feed-forward correction.** The addition of a delay line and Pockels cells allows Bob to store his photon until he receives the result of Alice's measurement and quickly make the corrective unitary transformation based on her message.

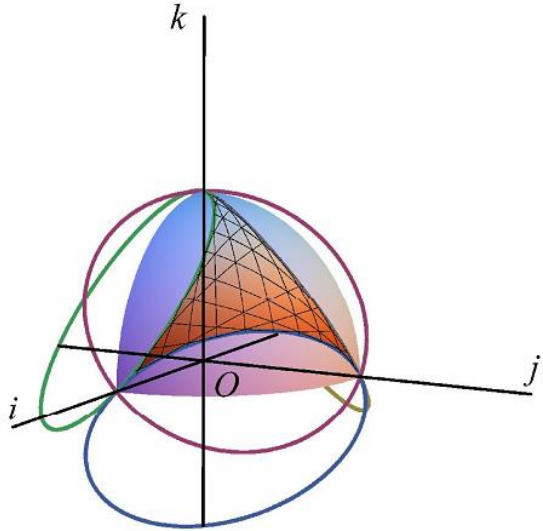

**Supplementary Figure 2 | A visual representation of measurement outcomes on equimodular states.** The orange triangular shaded area is the set of points giving the measurement outcome probabilities accessible by measuring equimodular qutrits in a basis mutually unbiased to the basis in which the two phases are applied. This area covers part of the positive octant on the unit sphere (the blue-purple-pink region) where vector *i*, *j*, and *k* correspond to the three orthogonal measurement outputs, where probabilities must be normalized (i.e., $i^2 + j^2 + k^2 = 1$). The colored circles geometrically define the shaded region. The blue, green, and yellow circles are pairwise tangent to each other at the points (1,0,0), (0,1,0), and (0,0,1), defining a fourth circle (shown in purple). These form a one-parameter subset of all accessible outputs which define the boarder of the shaded region. By plotting only the absolute value of these amplitudes, we find the part of the octant inside the shading. This region can be geometrically shown to encompass ~31% of all possible count rate combinations (i.e., 31% of the octant i, j, and k>0), using the circles shown on this diagram.

**Acknowledgements**
We thank Daniel Cavalcanti for helpful discussions. This work was supported by NSF Grant No. PHY-0903865, the NASA NIAC Program and NASA Grant No. NNX13AP35A. Partially supported by National Science Foundation Grants DMS-1201886, No. PHY 1314748 and No. PHY 1333903.

**Author contributions**
H.J.B. proposed the SDT protocol and with T.M.G. and P.G.K. devised the experimental implementation. T.M.G. performed the experiment and analyzed the results. M.J. led the packing number and volume calculations with T.-C.W.; T.M.G and T.-C.W. performed the Methods fidelity analysis; and H.J.B. performed the Supplemental Information measurement outcome analysis. All authors contributed to the manuscript preparation.